\documentstyle[11pt,newpaspman,twoside]{article}
\def\edcomment#1{\iffalse\marginpar{\raggedright\sl#1\/}\else\relax\fi}
\reversemarginpar

\begin{document}

\title{The Search for Terrestrial Planets: What Do we Need to Know?}

\author{Charles Beichman}
\affil{Jet Propulsion Laboratory,
California Institute of Technology,
Pasadena, CA 91109}

\begin{abstract}
The goal of finding and characterizing habitable planets in other 
solar systems represents one of humanity's greatest scientific 
challenges. NASA and ESA have 
initiated studies of missions that could accomplish this goal 
within the next ten years. What precursor knowledge do we need 
before we can initiate such a mission? How large should the first 
steps be in a program whose ultimate aim is to detect life on 
other planets? This talk describes different concepts for NASA's 
Terrestrial Planet Finder and discusses potential 
precursors in a program that balances scientific return, 
technological advance, and programmatic risk.
\end{abstract}

\section{Introduction}

The public and the scientific response to NASA's search for habitable 
planets and life has been very enthusiastic.  Most recently, the National 
Academy of Science's Decadal review  of Astronomy and Astrophysics (2000) 
endorsed the Origins goals,  noting:

\begin{quote}
``Key problems that are particularly ripe for advances 
in the coming decade\dots [include] studying the formation of stars 
and their planetary systems, and the birth and evolution of giant 
and terrestrial planets."
\end{quote}

\begin{quote}
``Search for Life outside of earth and, if it is found, determine 
its nature and its distribution in the galaxy\dots [This] is so 
challenging and of such importance that it could occupy astronomers 
for the foreseeable future."
\end{quote}

But the NAS panel added a caveat to its recommendation that NASA 
proceed with the Terrestrial Planet Finder mission (TPF):

\begin{quote}
``The committee's recommendation of this mission is predicated 
on the assumptions that TPF will revolutionize major areas of 
both planetary and non-planetary science, and that, prior to 
the start of TPF, ground- and space-based searches will confirm 
the expectation that terrestrial planets are common around solar 
type stars."
\end{quote}

This talk addresses what we must learn about the prevalence 
(or absence) of Earth-like planets before starting TPF and 
discusses what missions might play a role in a balanced program 
to advance  our scientific knowledge and technological prowess 
to the level  needed for TPF.

\section{Finding Planets Before TPF}

As described by the TPF Science Working Group, the primary challenges 
of the Terrestrial Planet Finder (TPF) are to ``\textit{detect radiation 
from Earth-like planets in the habitable zones} surrounding a 
statistically valid sample of $\sim$ 150 solar type (spectral 
types F, G, and K) stars, to characterize the \textit{orbital} and \textit{physical} 
properties of all detected planets to assess their habitability, 
and to characterize the \textit{atmospheres} and search for potential \textit{biomarkers} 
among the brightest candidates.'' The specified number of stars 
is a compromise between providing a high probability of success 
in finding habitable planets, setting a meaningful limit on the 
prevalence of planets in the event of a negative result, and 
the demands on the observing system.

The average distance to the stars with terrestrial planets is 
perhaps the most critical parameter that sets the scale of TPF. 
The baseline TPF concept calls for operation toward stars as 
far away as $\sim$15 pc, beyond which distance either the 
angular resolution or sensitivity become insufficient.
If, however, Earths were known to 
be ubiquitous, then one could concentrate on the closest stars 
(\texttt{<}5 pc) using smaller telescopes and lower angular resolution. 
Conversely, if Earths were known to be rare, it would be necessary 
to use larger telescopes to look beyond 15 pc to find even a few examples. 
Thus a major goal for a precursor TPF science program 
is to assess the fraction of stars with terrestrial planets 
in the habitable zones, denoted $\eta_\oplus$. A lesser but 
still valuable goal would be to assess the fraction of systems 
with giant planets located \textit{only} beyond the habitable zone 
so that Earths would have a chance to form and survive on stable 
orbits. 

\subsection{(Almost) Everything I Know About Planets I Learned from Radial 
Velocities}

As described at this conference, the radial velocity 
(RV) technique (Mayor and Queloz 1995; Marcy and Butler 2000) 
has proven to be the most successful technique to date in finding 
planets and even planetary systems. With almost 50 planets now 
known (see the invaluable \textit{Extrasolar Planets Encyclopaedia} at 
\textit{http://www.obspm.fr/planets}; Schneider 1996), we have learned that
approximately 5\% of stars have gas giant planets located within a few 
AU of their parent  star and ranging in size from 0.17/{\it sin\ i} M$_{J}$ 
(Marcy \&  Butler, this  conference; and Udry, Mayor \&
Queloz 2000) to beyond the 13 M$_{J}$ deuterium burning limit.
As many have noted, these gas giants 
all exist at orbital radii well within water-ice condensation 
line thought to be the interior limit for the accretion of such 
planets. Whether this discrepancy is due to dynamical effects 
or to an insufficient understanding of the mechanisms of planet 
formation is a matter of great debate.

While only two planetary \textit{systems} are known, ($\upsilon$ 
And and HD83443), Marcy (this conference) claims that the residuals 
in the RV data for half of his sample exceed the experimental 
uncertainties and are highly suggestive of the presence of planets 
with periods longer than the present observational datasets. 
While multiple systems may eventually prove be common, as yet 
we know of no counterpart to our own ``grand design'' solar system 
with gas giants on circular orbits located beyond the water-ice 
condensation line and rocky planets nestled within the habitable 
zone. Furthermore, the broad range of eccentricities and small 
orbital radii of the known giant planets may be inconsistent 
with the existence of habitable, terrestrial planets. Some have argued that 
these unexpected results mean that solar systems like our own 
are rare. However, many would argue that we are simply reaping 
the fruits of using a technique (radial velocity) that is biased 
toward finding gas giant planets on short period orbits. What 
are the properties of the solar systems (if any) orbiting the 
95\% of solar type stars for which the RV technique has to date 
set only upper limits? To make further progress we need RV data 
over longer periods to probe larger orbital distances, 
but we also need qualitatively new types of data.

\subsection{Transit Experiments}

A second technique that has recently borne fruit is the measurement 
of the transit of a planet in an edge-on orbit across the face 
of its parent star. The detection of the transits of HD 209458 
(Charbonneau et al.  2000; Henry et al.  2000) is a spectacular 
demonstration of the power of this technique and a validation 
of the planetary interpretation of the RV data. In addition to
giving the planet's true mass (by determining the inclination angle),
radius and density, spectroscopic observations might even reveal information
the composition (Sasselov and Seager 2000). On 
the other hand, the failure to find any transits toward to 3.4$\times10^{4}$ 
stars in the globular cluster 47 Tucanae (Gilliland et al.  
2000) is an indication that not all environments are hospitable 
to the formation or survival of planets. 

While Jupiter-sized planets present a readily observable transit 
signal of 1.5\%, Earths produce a signal that is only $\sim 10^{-4}$ 
in depth. However, if the $5\times 10^{-5}$ accuracy of the recent HST 
observations of HD 209458 (Brown et al.  2000, this conference) 
can be replicated or improved on in an imaging system, then the 
detection of Earths around other stars is plausible. A wide area 
survey of $\sim10^5$ stars is required because transits are 
rare due to the requirement for precise alignment 
--- 1 (10)\% for planets at 1 (0.1) AU --- and infrequent due 
to orbital periods ranging from a few 
months up to once per year for a planet at 1 AU. 

A number of groups are either already making such observations 
on the ground (Borucki et al.  2000; Howell et al.  2000) 
or proposing them for space missions, including Kepler in the 
US (Koch et al.  1998) and Eddington/COROT in Europe (Leger \textit{et 
al}. 2000). These projects would target solar type stars typically 
a few 100 to 1,000 pc away in fields of 10-100 sq. deg. Since 
these weak signals must be measured at least three times to 
identify and then confirm their presence, a mission dureation of 
at least 3-4 years is required to find planets in the habitable zone. 

\subsection{Microlensing Searches}

Searches for planets by gravitational microlensing take advantage 
of the fact that the presence of a second mass in a lens system 
produces multiple images of a background star. The lensing star 
produces a magnification of the background star's brightness 
that lasts a month or longer and which can be as great as a factor 
of 3 to 10. The second mass, located within a few AU of the central 
star, produces an additional magnification of a few 
percent up to a factor of two, lasting anywhere from a few days 
for a Jupiter to a few hours for an Earth (Peale 1997). A number 
of groups have reported detections of binary stars and, with less 
certainty, of objects of Jovian mass (Alcock et al.  2000). 
The frequency of these events is, unfortunately, quite low so 
that $\sim 10^{8}$ stars must be monitored continuously if the 
hours-long enhancements due to an Earth are to be detected. 

Advantages of the microlensing technique include the large photometric 
signal, the early identification of the lensing system which 
can then be constantly monitored, the well defined signature 
of a microlensing event, and the ability to find free floating 
planets. The disadvantages include the great distances to the 
lensing star and the background stars (4 and 8 kpc, respectively), 
and the fact that the derived mass of the planet depends on the 
poorly known stellar mass. Further, most of the lensing stars 
will typically be M dwarfs or smaller and thus not direct solar analogs.
Proponents of both ground-based programs and space-based 
missions such as GEST (Bennett et al.  2000) have suggested 
that microlensing can reliably find  tens of Earths per year. 

\subsection{The Promise of Astrometry}

The complementary technique to radial velocities for searching 
for planets around \textit{nearby stars} is astrometry. Astrometry 
is relatively unaffected by the photospheric motions that limit 
the RV technique to detections of $\sim$ Saturn-mass planets.
Furthermore, astrometric signals grow rather than 
shrink with increasing orbital distance, reversing the selection 
bias that affects RV measurements. The scale of the astrometric 
measurement challenge is set by two cases: the signal of a Jupiter 
orbiting 5 AU away from a G2V star 10 pc away is $\sim$500 $\mu$as 
while the astrometric signal of an Earth orbiting 1 AU away the 
same star is $\sim$0.3 $\mu$as. Detecting a planet of unknown 
period and amplitude  requires a dataset extending at least as 
long as half the planet's orbital period. For reliable planet detection, 
the \textit{single measurement accuracy} must be approximately half of 
the astrometric amplitude of the planetary signal (Brown et al. 1999).

\begin{table}
\centering
\caption{ \it Table 1. Prospects for Astrometric Detection of Planets}
\vskip 0.1 in
\begin{tabular}{|l|r|r|r|r|r|r|} 
\hline
& 		& 		& Single 	& 	 & Dist to &Dist to\\
& 		& Largest	& Meas. 	&Limit.& Detect  &Detect\\
&Start/End	&Orbit 	&Accur 	&Mag.  & Jupiter & Earth\\
Project&Dates & (AU)    &($\mu$as) 	&(mag) & (pc)    &    (pc)\\
\hline
Keck-I&2003-&15& 30 &15 &85&---\\
\& VLT-I &2025 & & & & & \\
FAME	&2005-2010&5&500&9&5&---\\
SIM 	&2008-2013&5&1-4&10&$>$1,000&3\\
GAIA  &2012-2017&5&40&10&60&---\\
\hline
\end{tabular}
\end{table}

As summarized in Table 1, the two NASA projects
for astrometric planet detection are the Keck Interferometer 
and the Space Interferometer Mission (SIM; Unwin et al. 1998). 
Keck-I will be capable of detecting Jupiters out to almost 100 pc
and Uranus-mass planets at 5 AU out to $\sim$5 pc. And 
since it will operate for up to 25 years, Keck-I 
(and eventually VLT-I in the southern hemisphere) will be able 
to find planets on long period orbits and determine their masses. During
its 5 year mission SIM will be able to detect planets of just a 
few Earth masses in 1-5 AU orbits around 
stars as far away as 10 pc. SIM will lower the mass limits into the 
range predicted for the ``rocky'' as opposed to ``gas giant'' planets.

There are two other astrometric missions worth mentioning in 
this context. NASA is planning to launch the full Sky Astrometric 
Explorer (FAME) around 2005. ESA has recently selected the GAIA 
mission as a Cornerstone for a launch around 2012. It
is important to recognize that these survey 
missions accomplish their final mission accuracies after averaging 
$\sim$1000 individual measurements. The \textit{single measurement 
accuracy} which is critical for planet detection is $\sqrt 1000\sim 30$ 
times worse than the \textit{mission} accuracy, or 1,500 $\mu$as for 
FAME and 120 $\mu$as for GAIA. While averaging of the 
data can improve the effective single measurement accuracy (say 
by a factor of 3), FAME and GAIA will find only giant planets.

\section{The Challenge of Direct Detection}

Finding habitable environments or even extant life ultimately 
requires the direct detection of photons from planets. 
The  challenges include faint signals, an enormous contrast ratio 
between the star and the planet, the close proximity of the planet 
to the star (1 AU corresponds to 0.1$^{\prime\prime}$ at 10 pc), and the 
presence of zodiacal emission in our own and in the target solar 
systems. But as described in the ExNPs Roadmap, 
the Terrestrial Planet Finder monograph (Beichman, Woolf 
and Lindensmith 1999), and the recent Darwin report, a mission capable
of achieving this goal is within our grasp.  

A broad suite of TPF architectures is being investigated by 
four NASA-sponsored studies involving 16 industrial 
concerns, 30 universities, 75 scientists, including a number 
of European researchers.  The major concepts under investigation
include nulling IR interferometers, visible light telescopes 
with coronagraphs or apodized pupils, and
sparse aperture telescopes. The TPF project will choose 4-8 of 
the most promising concepts for more detailed study 
over the next 18 months. 

\subsection{Reflected Light Systems}

Observing systems to detect the 
reflected light from an Earth 10 pc away must utilize a large 
visible light telescope ($\sim$50-100 m$^{2}$ of collecting area) 
with an advanced coronagraph and/or  pupil apodization
along with precise wave front control ($<\lambda/1000$). 
The advantages of these systems include operation in a traditional 
imaging mode on a single spacecraft. The chief disadvantages 
arise from the extreme star:planet contrast ratio (\texttt{>}10$^{9}$) 
which implies the need for exquisite suppression of
diffracted and scattered light on a large (monolithic) telescope. Spectral 
features of H$_{2}$O and O$_{2}$ might be used to characterize planetary 
atmospheres and look for life. What makes visible coronagraphs 
interesting again after a decade of neglect is the new technology 
of small deformable mirrors with thousands of actuators capable 
of \texttt{<} 1 nm stability. The combination of a large telescope 
of diameter, D, a coronagraph, and a deformable mirror can, in 
principle at least, reject scattered and diffracted starlight 
in a field of a few arcseconds around a bright star (Malbet, 
Yu and Shao 1995). The number of actuators required is given 
by $[Field/(\lambda/D)]^{2}$ or roughly
$[3^{\prime\prime}/(0.25^{\prime\prime} /(0.5 \ \mu m /8\  m)]^{2} \sim$ 
37,000 for a 3$^{\prime\prime}$ field on an 8 m telescope. While the issues 
of telescope quality, stability, control of all sources of 
scattered light, etc., make the coronagraph approach very challenging, 
the technique is under active study. \\

\subsection{Thermal Infrared Systems}

Observing systems to detect thermal radiation incorporate a 
nulling infrared interferometer (Angel and Woolf 1997; 
TPF and Darwin reports) consisting 
of 4-6 telescopes, each 2 m (Darwin) - 3.5 m (TPF) in diameter, 
operating on separate spacecraft over baselines of 75-200 m 
to achieve the required angular resolution. 
The chief advantages of these systems include a relatively favorable 
star:planet contrast ratio (10$^{6}$) and the presence of broad, 
deep absorption bands that can characterize planetary 
atmospheres, e.g. CO$_{2}$, H$_{2}$O, or serve as signposts 
of life, e.g. O$_{3}$ and CH$_4$.  The disadvantages of the interferometers
include noise due to emission from zodiacal dust, the
need for multiple spacecraft, and cryogenic operation.

\section{TPF Precursors}
Detecting gas giant planets is obviously easier than detecting 
Earths. For example, a 2 m telescope equipped with a coronagraph 
could detect and characterize Jupiters at visible wavelengths 
(the ECLIPSE mission; Trauger et al.  2000). Similarly, a 
smaller version of the infrared nulling interferometer could 
investigate the spectra of gas giant planets and even detect 
Earths around the closest stars. For example, the observing time, $\tau$, 
for a nulling interferometer to detect an Earth 10 pc away at 12 $\mu$m scales roughly 
as follows (Beichman and Velusamy 1999):

$$ \tau \sim  (2\  hr)  \times 
	( { {R_{Planet}} \over {R_\oplus} } )^{-4} \times
	( { {Tel.\ Diam.}    \over {3.5\ m     } } )^{-4} \times 
	( { {Dist.    }    \over {10\ pc     } } )^{4} \times
 	( { {Resln   }    \over {3         } } ) \times 
	( { {SNR     }    \over {5         } } )^2 $$

Planet diameter, telescope diameter, and distance to the planet 
can all be traded against one another linearly. Thus, a scaled 
down version of the nulling IR interferometer, e.g. TPF-Lite 
using four 1.5 m telescopes on a fixed 20 m boom, could carry out 
spectroscopy of Jupiters and even detect an Earth around the 
stars within 5 pc (Woolf et al.  1998; Velusamy and Beichman 
2000). This system would require relatively little new technology 
beyond the nulling capabilities being developed for the Keck 
Interferometer.

These modest-class missions, operating at either visible or infrared 
wavelengths, would give us valuable information on the physical 
properties of some of the gas giant planets already detected 
(Sudarsky et al.  2000), would identify new gas giants on distant 
orbits that would take many years to detect via astrometric or 
radial velocity measurements, and would give us real world experience 
with complex new observing techniques. 

\begin{table}
\centering
\caption{ \it Table 2. Summary of Ground-Based Capabilities }
\vskip 0.1 in
\begin{tabular}{|l|r|r|r|r|r|r|} \hline
    		&Tel.     &Time to 	&	&Planet &Dist to & \\
Technique	&Size     &Return   	&Planet  &Orbit&Planet& \\
    		&(m)      &	Data (yr)   &Size	& (AU)& (pc) &Spectra \\
\hline
Radial Velocity&3-10 &Now&$>$Saturn&$<$ 3 &$<$50&None\\
Astrometry     &2    &5&$>$Uranus&$<$15&$<$ 25&	None	\\
Transits       &1-2  &Now&$<$Uranus	&$<$ 2&500-1,000&None\\
MicroLensing   &2-3  &Now&Saturn &	1-5& 4,000&None\\
		   &     &	 & Earth? &    &      & \\
Interferom. &8-10 &2&Jupiter&$<$1&$<$25 &Yes\\
Imaging     &     &	&	  &    &      & \\
\hline
\end{tabular}
\end{table}

\section{What Can (Must) We Know Before Starting TPF?}

How will our scientific knowledge increase over the next decade 
to satisfy the requirements laid down by the NAS? Tables 2 and 3 
summarize the information different facilities 
will (or could) provide toward answering the NAS concerns. Relevant
portions of the on-going NASA progam include the following:\\


$\bullet$ The Keck Interferometer and the European VLT-I will use astrometry 
to extend the present (radial velocity) census of giant planets 
over wide range of orbital parameters. With these data in hand, 
we can use dynamical and evolutionary theories to infer a great 
deal about the prevalence of terrestrial planets, and the suitability 
of particular stars as potential TPF targets.

$\bullet$ Early in the next decade SIM will push the census 
of planets around nearby stars down to masses of terrestrial 
planets, highlighting good targets for TPF. Note that of all 
the proposed new capabilities, only SIM measures the masses, 
information that is essential to understanding the habitability 
of solar systems and of planets themselves.

$\bullet$  SIRTF, Keck Interferometer, and the Large Binocular Telescope 
(LBT) will characterize zodiacal dust clouds around nearby stars 
to a level of \texttt{<}10 times that of the solar system. We will 
be able to extrapolate the ``disk luminosity function'' to near 
Solar System levels to assess the importance of this noise source 
on the problem of planet detection.\\


\begin{table}
\centering
\caption{ \it Table 3. Summary of Spaced-Based Capabilities}
\vskip 0.1 in
\begin{tabular}{|l|r|r|r|r|r|r|} \hline
    		& Tel.& Time to 	&	&Planet &Dist to & \\
Technique	& Size& Return   	&Planet&Orbit&Planet& \\
Project    	&  (m)& Data (yr) &Size	& (AU)& (pc) & Spectra\\
\hline
SIM 		&0.4	&5-10 & $>$Few Earth&$<$5&$<$10& None\\
Transits 	&1-2 	&7-9 	&Earth &$<$2&1,000&Limited\\
         	& 	& 	&(FGK stars) & & & \\
MicroLensing&1-2 	&	5-7 &Earth&1-5&4,000&None\\
         	& 	& 	&(M stars) & & & \\
\hline
\hfill TPF-Lite & & & & & &\\
Interferom. & 1-2 &8&Saturn &$<$5&$<$25&Atmosphere\\
		& 1-2 &8&Earth &$<$5&$<$5&None\\
Coronagraph &2 	& 8& Jupiter&$<$5& $<$10 &Atmosphere\\
\hline
\hfill TPF 	& 	& & & & &\\
Interferom. & 3.5	& 10-15 &Earth&	$<$5&	$<$15& Life\\
Coronagraph & 6-10& 10-15 &Earth&$<$5&$<$15& Life\\
\hline
\end{tabular}
\end{table}

But these projects do not provide all the information we might 
require. Projects that are not presently in the program that would be of 
considerable utility include the following:\\


$\bullet$ A mission to determine $\eta_\oplus$ from either transits or 
microlensing would set the distance scale for 
TPF and determine the required aperture size and angular 
resolution needed for a high probability of success in finding 
Earths. While the COROT mission is a good first step, a mission
on the scale of Eddington or Kepler is required for robust statistics.
An intensive ground-based microlensing campaign would a valuable near-term adjunct to
this effort.

$\bullet$  An advanced coronagraphic telescope might detect and characterize 
giant planets around nearby stars and demonstrate the technologies 
needed for more capable future missions.

$\bullet$ A scaled down version of a nulling IR interferometer would detect 
and characterize giant planets around nearby stars, detect 
Earths (if any) around the nearest stars, and demonstrate 
the technologies needed for more capable missions in the future.\\


From the programmatic standpoint, the
problem with these intermediate missions is that they will 
be expensive and are likely to delay the start of TPF,
particularly if we have  to wait for their results. The transit, microlensing, and  
coronagraphic missions have all been proposed for NASA's Discovery 
program which has a $\sim$\$300M cost cap; realistic missions 
might cost even more. A scaled down nulling IR interferometer 
could easily cost as much as SIM. Thus if we require
these missions as precursors, we must accept 
delays in TPF and higher program costs. 

Consider how we might combine the following four elements into 
a program:

\textit{Element 1. Initiate a Mission to Determine $\eta_{Earth}$ 
(Launch in 2006).} If an adequate number of stars can be observed 
with the requisite precision in a Discovery-class mission, then 
either a transit or a microlensing experiment would bolster our confidence 
in the goals of TPF and set the distance scale for planet searches. 

\textit{Element 2. Initiate a TPF-Lite Mission (Launch in 2008).} An 
IR nulling system could make spectroscopic studies of already 
known giant planets, survey for new giant planets on distant orbits, 
and even survey the nearest stars for Earths. A visible light coronagraph like ECLIPSE 
could target Jupiters around nearby stars.

\textit{Element 3. Initiate Nominal TPF in 2007 (Launch in 2012).} The 
present NASA plan for TPF involves an IR nulling interferometer 
based on technology from the SIM, NGST, and ST-3 missions. With 
a suitable redirection of technology funds, this plan would support 
a coronagraph alternative if the results of the ongoing studies 
are favorable. However, this schedule would not allow us to incorporate 
into the \textit{design} of TPF the knowledge gleaned from a transit 
or microlensing experiment, or from the SIM mission. The scale 
of TPF would have to be set by the compromise described above 
of wanting to observe $\sim$150 stars out to $\sim$15 pc. 
TPF would however have a well characterized list of targets from 
radial velocity observations and astrometry using Keck-I,
VLT-I, and SIM. 

\textit{Element 4. Launch TPF based on knowledge of prevalence of 
Earths (Launch after 2015).} If it is necessary to {\it design} TPF 
based on the known incidence of Earths and the actual configuration 
of (giant) planets around nearby stars, then we must wait 
until data are obtained from the statistical experiments 
(transits or microlensing) and/or results from SIM. Under this scenario, 
TPF could not be launched before 2015.  Achieving a high degree of 
certainty requires that we wait a very long time! 

The most aggressive approach, i.e. the present Origins plan,
would be to proceed at full speed with the nominal TPF (Element 
3). While one never has absolute certainty of what one will find 
in exploration, \textit{a TPF capable of looking for Earths 
around a few hundred stars out to 15 pc has a high probability 
of success except for the most pessimistic estimates of} $\eta_{Earth}.$ 
A more deliberate approach would require an early measurement 
of $\eta_{Earth}$ (Element 1), would fly a modest-scale precursor 
(IR interferometer or coronagraph, Element 2), and would finish 
with a very capable spectroscopic mission in the middle of the 
next decade (Element 4).  

The decision between the two approaches should be made on the 
basis of technological readiness. If the development of 
the necessary TPF technologies proceeds
smoothly, then we should move as swiftly as budgets will
allow to the nominal TPF. If not, then  a more deliberate approach makes 
sense, allowing near-term discoveries and demonstrations of complex 
observing systems in space. Over the next year, NASA must assess these options 
and in conjunction with ESA and other international partners 
come up with a viable program that addresses the age-old questions 
that have motivated this most exciting conference.

{\bf Acknowledgements.} This work was carried out for NASA by JPL under a contract with 
the California Institute of Technology.

\end{document}